\documentclass{jcgt}
\usepackage{booktabs}
\usepackage{xcolor}
\usepackage[T1]{fontenc}
\usepackage{tgtermes}
\usepackage{tgheros}
\usepackage{courier}

\setciteauthor{Daniel Szelogowski}
\setcitetitle{Temporal Persistence for Stylized Rendering: Historical Image Storage and Reconstruction}
\setheadtitle{Temporal Persistence for Stylized Rendering}
\begin{document}

\title{Temporal Persistence for Stylized Rendering:\\Historical Image Storage and Reconstruction}
\author{Daniel Szelogowski~\href{https://orcid.org/0000-0002-0350-5771}{\includegraphics[width=8pt]{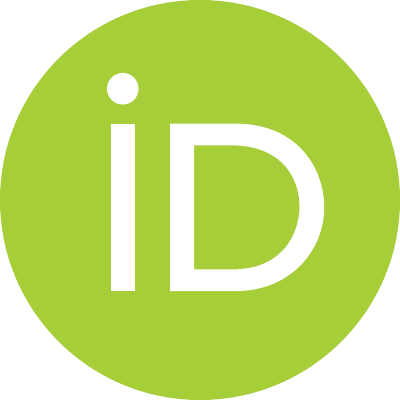}}\\University of Wisconsin--Whitewater}

\teaser{\includegraphics[width=\linewidth]{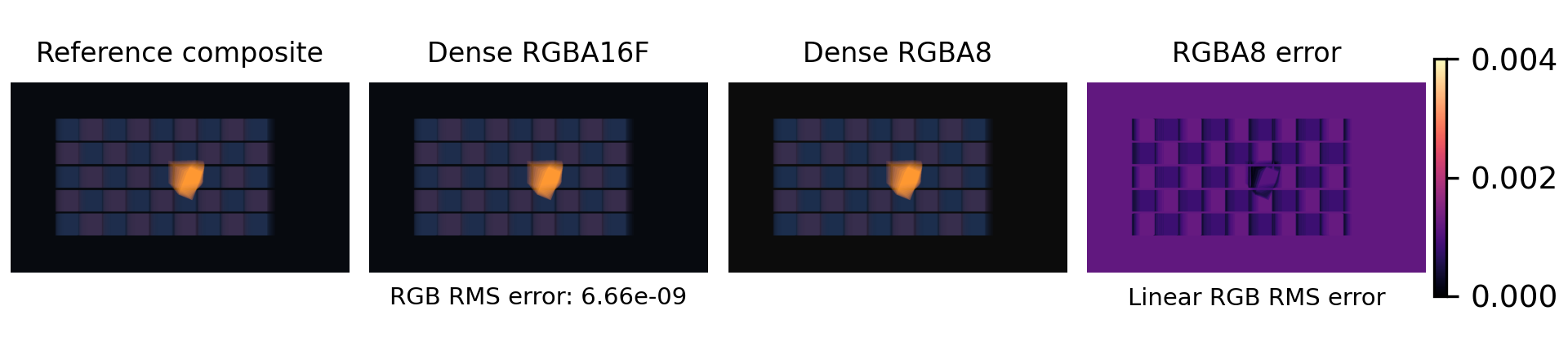}
\caption{A moving camera produces overlapping historical silhouettes in a procedural scene. The first three panels show the reference composite and two dense storage formats; the fourth shows RGBA8 error on a fixed linear RGB scale. The half-float panel reports its numerical error beneath the image, where a shared-scale error map would appear black.}
\label{fig:teaser}}

\maketitle

\begin{abstract}\small
The overlapping silhouettes and gradual fading in \emph{The Legend of Zelda: Majora's Mask}\linebreak (released April 27, 2000) produce a distinctive depiction of motion blur, in which earlier scene states remain visible as part of the current image. Reproducing this appearance in a modern renderer involves decisions about when historical images are sampled, how their contributions are retained, and how the resulting history can be represented within the available memory and processing budget. In this paper, we investigate these decisions through complete image histories, shared image regions, and selectively retained pixel changes, using Godot engine implementations evaluated against independently composited references. For a ten-image history sampled at 20 Hz on an NVIDIA GeForce RTX 3060 Ti, half-float storage halves history allocation and reduces median storage-pass duration by factors of 2.40 at 720p and 1.73 at 1080p, preserving the supplied source precision. The adaptive experiments examine how memory savings depend on reconstruction metadata and allocation, including a sparse representation that reduces allocation by 83.81\% at low resolution with 445.25 times the GPU duration.
\end{abstract}

\section{Introduction}
In \emph{The Legend of Zelda: Majora's Mask}, overlapping silhouettes preserve earlier scene states as they progressively fade, producing a distinctive layered depiction of motion. Its layered appearance makes the positions, shapes, and relative prominence of historical images part of the visible effect, with the spacing between silhouettes conveying successive moments of motion with a dream-like effect. This appearance motivated us to investigate how a modern renderer could retain control over those historical contributions as image resolution, sampling, and storage requirements change. Temporal persistence describes this continued visibility of earlier images, and a framebuffer --- the image storage into which a renderer writes its output --- provides one means of retaining the information needed to construct it. We study the resulting relationship between silhouette appearance and historical image representation through an explicitly defined experimental target in Section~2. Our evaluation examines image fidelity, allocated memory, and the time required to update and combine that history on a \textbf{graphics processing unit (GPU)}.

\begin{figure}[htbp]
\centering
\begin{tabular}{cc}
\includegraphics[width=0.47\linewidth,height=1.2in,keepaspectratio]{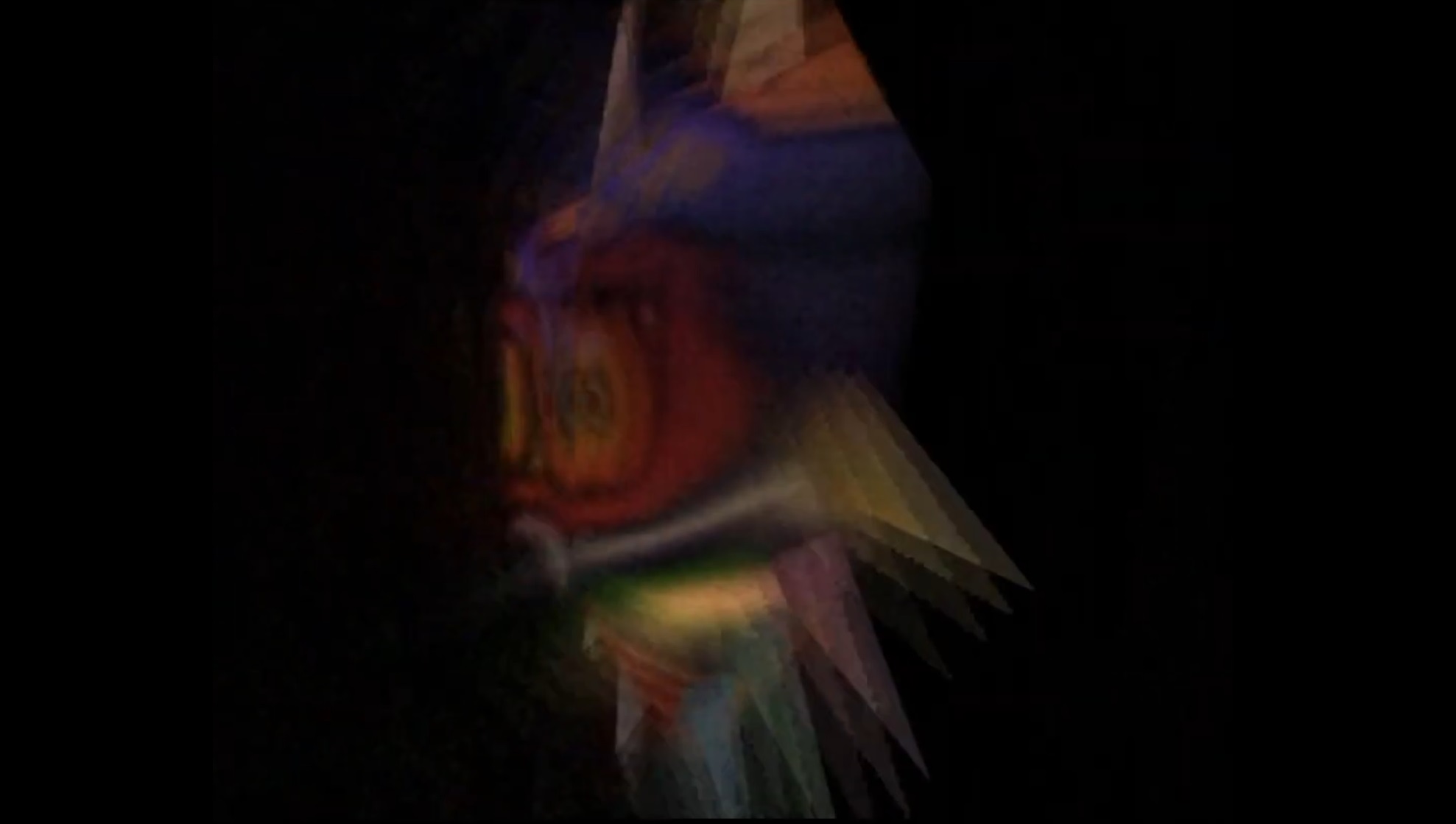} &
\includegraphics[width=0.47\linewidth,height=1.2in,keepaspectratio]{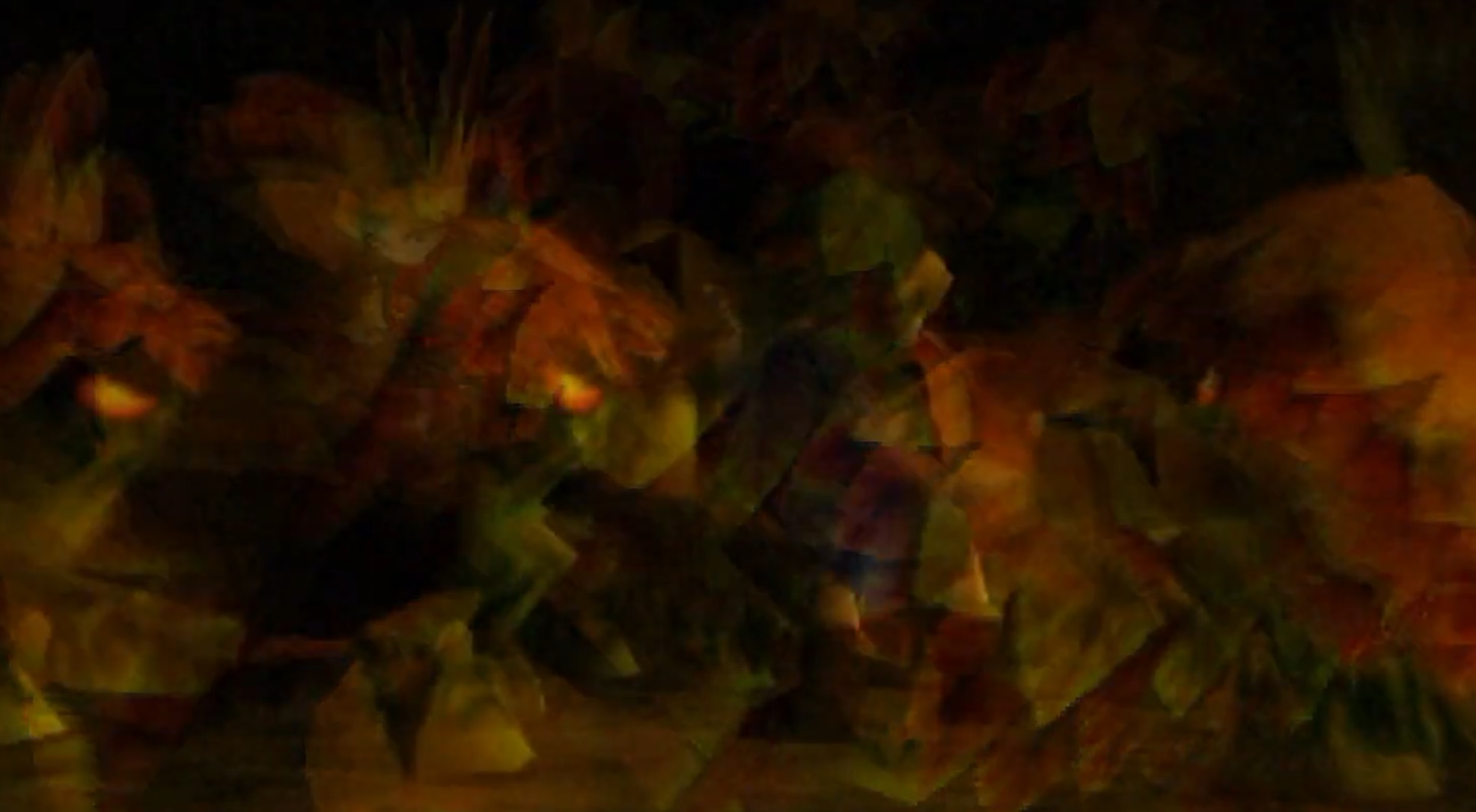}\\
\small (a) Overlapping mask silhouettes & \small (b) Overlapping scene figures\\[5pt]
\includegraphics[width=0.47\linewidth,height=1.2in,keepaspectratio]{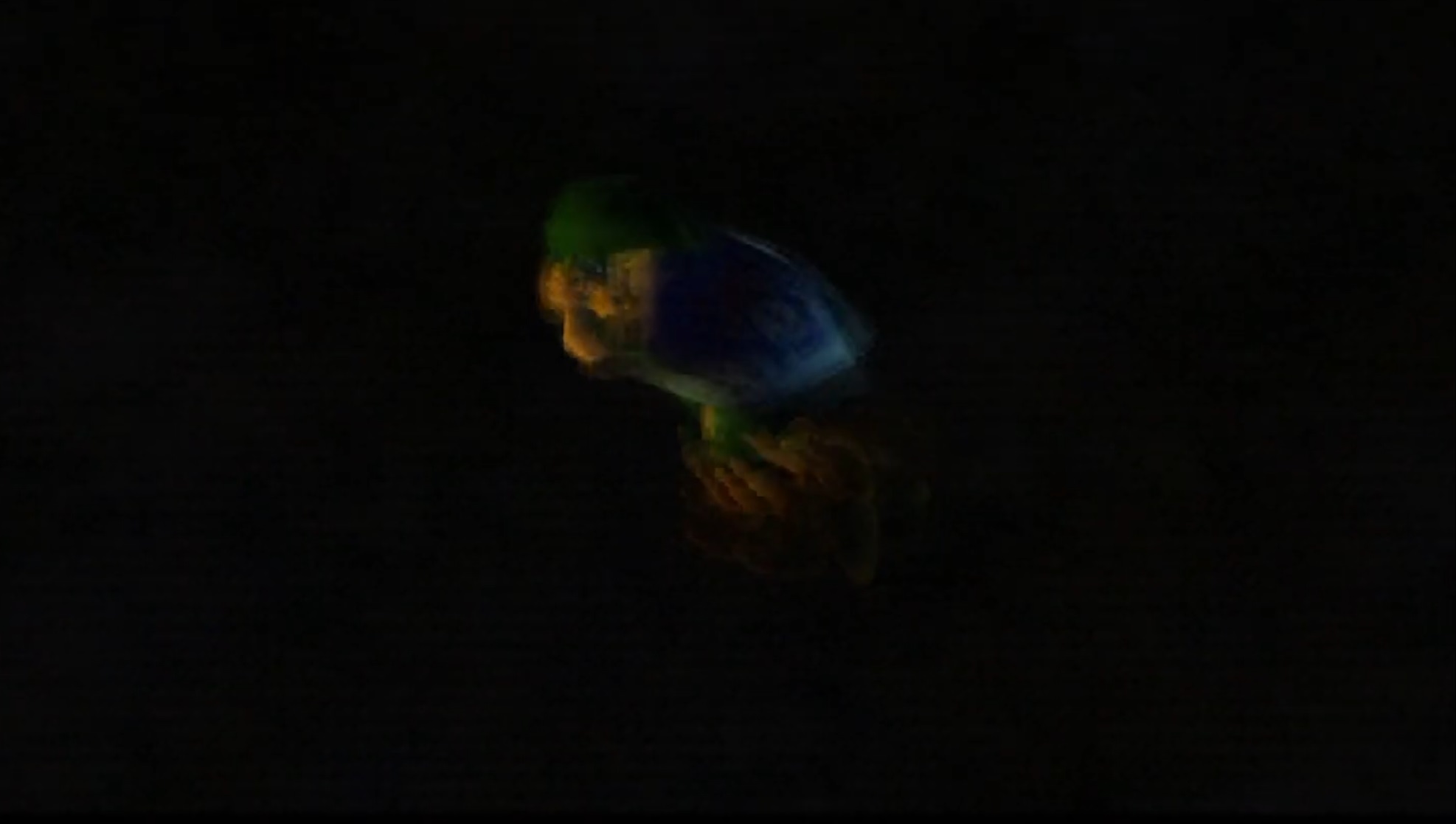} &
\includegraphics[width=0.47\linewidth,height=1.2in,keepaspectratio]{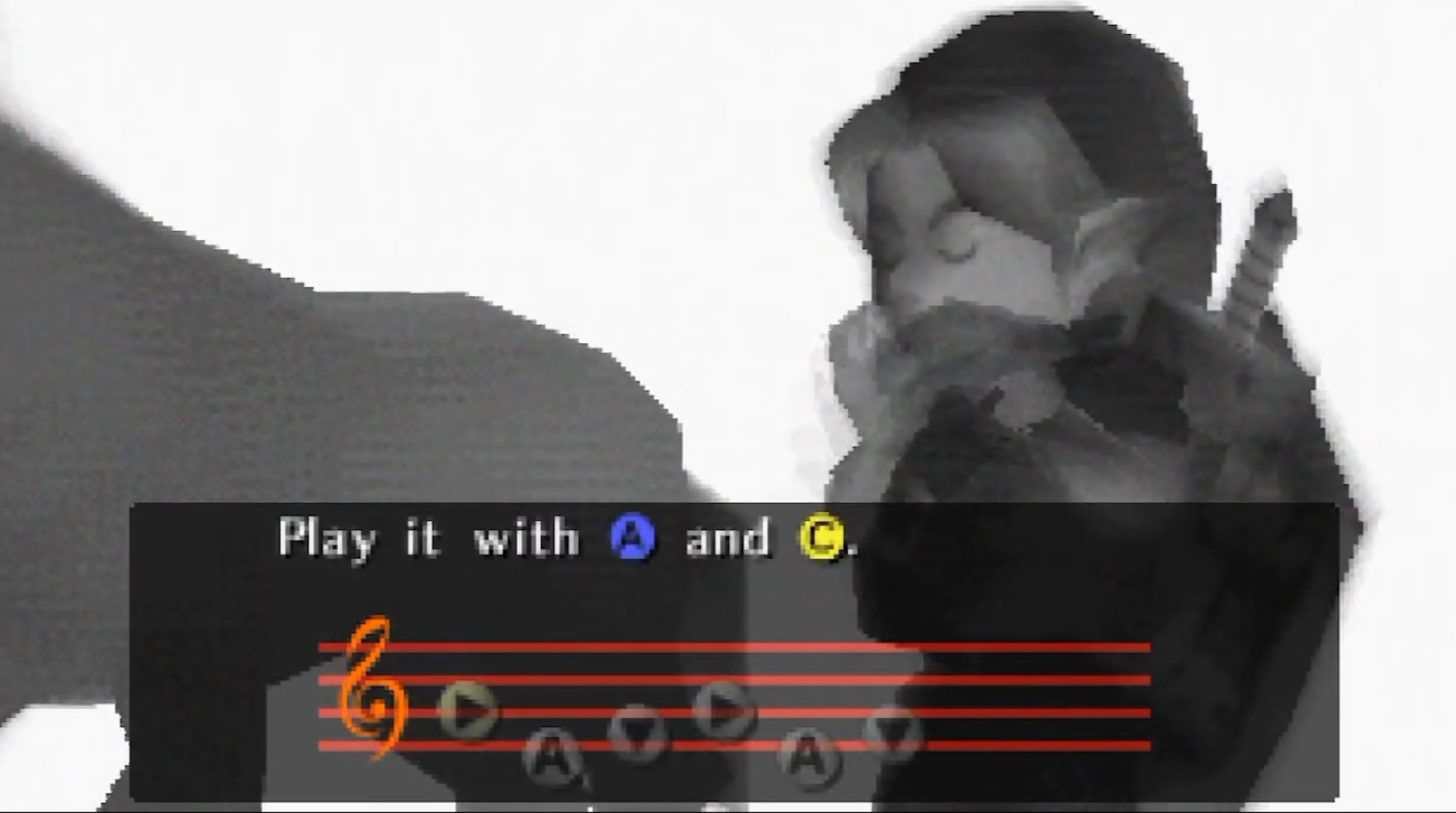}\\
\small (c) Fading historical silhouette & \small (d) Scene persistence and interface
\end{tabular}
\caption{Reference examples from \emph{The Legend of Zelda: Majora's Mask}, showing the layered silhouettes that motivated this study. Game imagery \copyright\ Nintendo.}
\label{fig:mm-reference}
\end{figure}

The examples in \underline{Figure~\ref{fig:mm-reference}} show how the effect preserves recognizable earlier shapes at different opacities. Overlap gives motion a layered appearance, while the fading of older contributions determines how long that history remains visible. The interface in panel~(d) also provides a visual example of the separation between the scene and the overlaid text. We use these images to illustrate the motivating appearance, with quantitative comparisons in this paper performed against the independently rendered experimental reference.

The simplest storage option is recursive feedback, which blends each incoming image with one accumulated image containing the contributions of earlier updates. Its memory requirement is small because the complete history is represented by that single composite. The retention coefficient determines how much of the previous composite survives each update, so a fixed coefficient produces a shorter trail when updates occur more frequently. Adjusting the coefficient according to elapsed time preserves the exponential decay duration, while historical silhouette placement continues to follow the supplied capture times. A finite image history provides direct access to each retained scene state and allows its contribution to be removed when a specified age is reached. We separate \textit{capture} --- the acquisition of a scene image, from \textit{presentation} --- the display of the resulting composite, and evaluate storage after the scheduled images have been acquired.

Our comparison begins with dense storage, which reserves a color value for every pixel at every retained timestamp. We then examine two adaptive representations: shared tiles, which reuse or combine rectangular image regions, and sparse innovations, which retain selected pixel changes between consecutive captures. Both seek to exploit temporal redundancy, meaning repeated or closely related information across successive scene images. Development experiments at low resolution allowed us to examine their approximation errors and account for the auxiliary data needed to locate retained values. We subsequently evaluated dense color formats at native rendering resolutions, where precision reduction offers a useful comparison with adaptive allocation. This sequence of experiments connects the mathematical representation to its actual GPU cost and identifies the conditions behind the observed memory savings and reconstruction failures.

\subsection{Framebuffer Persistence on the Nintendo 64}
The original implementation constructs persistence by reusing the rendered scene image through the Nintendo 64's framebuffer and blending hardware. The \textbf{Reality Display Processor (RDP)} provides fixed-function texture, color-combination, and blending operations, allowing an existing framebuffer to be drawn over newly rendered scene content. In the inspected US 1.0 decompilation of \emph{Majora's Mask}, the procedure \texttt{Play\_DrawMotionBlur} first enters a setup state that copies the current scene into a saved framebuffer; subsequent active updates blend the saved image over the current scene and then save the resulting composite \citep{MMDecompPinned}. Since this saved image already contains earlier contributions, each update carries their remaining weight forward through a \textbf{recursive feedback} loop. This accounts for the successive fading silhouettes using one accumulated image, with their weights and positions determined by the blend parameters and the scene states supplied at each update. Kaze Emanuar's explanation provides an accessible account of this sequence and its visual consequences \citep{KazeEmanuar2026}.

The mechanism uses the hardware's existing image operations, and its resource requirements follow from reading, blending, and copying framebuffer data. The inspected \texttt{PreRender} path uses 16-bit color storage, point-filtered framebuffer drawing, and a caller-supplied environment alpha, with additional dithering behavior in the translucent path \citep{MMDecompPinned}. At a conditional resolution of $320\times240$, one extra image with two bytes per pixel occupies 153,600 bytes --- approximately 150 KiB --- in addition to the renderer's existing buffers. An idealized update reads the saved and current images, writes the blend, and copies the result back to history, accounting for five image-sized color transfers or approximately $10P$ bytes for $P$ pixels. This is a theoretical traffic estimate; hardware caching, coverage information, command processing, and display output require additional treatment for an actual performance model. Emanuar's video estimates that the operation consumes more than one fifth of a frame's rendering budget remains an attributed estimate rather than a hardware timing measurement from this study \citep{KazeEmanuar2026}.

The placement of the copy also explains the use of a separate saved image alongside the display buffers. In the normal inspected draw path, scene blur and history capture precede the drawing of the \textbf{heads-up display (HUD)} and message text, allowing those interface elements to remain outside the accumulated image. Reading the final displayed framebuffer as history would also retain those later overlays, changing the appearance of subsequent composites. The video illustrates retention with an opacity of $180/255$, approximately $0.706$, so a seeded contribution retained twice has idealized weight $(180/255)^2\approx0.498$; the source routine permits callers to select the opacity, and fade schedules depend on its use. Its history is updated when the draw operation executes, whereas the independent 20 Hz schedule and ten-image finite window used in our experiments are explicit choices for studying sample placement and storage. Hence, the historical implementation explains both the motivating visual behavior and the importance of capture placement, while our evaluation examines representations with independently specified historical contributions.

\subsection{Related Work}
Programmable motion effects allow an artist to control how motion appears through the representation and combination of scene information across time \citep{Schmid2010}. Related work on cartoon motion cues examines how a rendered image can convey movement through deliberately constructed visual features \citep{Collomosse2005}. These approaches inform our treatment of \textbf{historical silhouettes} as visible elements whose placement and fading are part of the intended appearance. Motion blur introduces a related reconstruction problem through the integration of scene radiance over a camera exposure; image-based filters estimate this appearance from the information available to the renderer \citep{McGuire2012,Guertin2014}. The surveys of motion blur and temporal coherence provide a broader account of the sampling, visibility, and reuse decisions involved in combining images across time \citep{Navarro2011,Scherzer2012}. Temporal antialiasing applies historical reuse to estimating the current image \citep{Yang2020}; our experiments retain explicitly weighted earlier scene states so that their silhouettes remain visible throughout a finite history window.

Texture compression illustrates the relationship between a compact image representation and the work required to recover its values during rendering \citep{Beers1996}. For an adaptive temporal history, this relationship includes the addresses, shared references, and allocation decisions needed to locate retained information before composition. Event-camera reconstruction provides a related use of temporal intensity changes, including their integration into image estimates and their use in asynchronous image operations \citep{Scheerlinck2019Intensity,Scheerlinck2019Convolutions}. Our \textbf{sparse innovation} representation derives signed changes from successive complete scene images, assigning retention priorities according to their remaining contribution to the target composite. The shared-tile representation uses least squares to construct a representative image region from two retained regions, with its objective incorporating their future temporal weights. We evaluate these established approximation ideas through their resulting image error, full storage requirements, and GPU execution costs.

\section{Target and Presentation}
We define the target effect as a weighted average of scene images captured at regular intervals. Let $F_n=F(nT)$ denote the unmodified scene color at capture index $n$, with $T$ seconds between scheduled captures. Colors are represented in linear light, where weighted averaging operates on values proportional to scene light intensity. The history length $H$ is the maximum number of captured images contributing to an output, and the half-life $h$ is the time over which an unnormalized exponential weight decreases by half. With retention $a=2^{-T/h}$, the number of available samples $c_n$, normalization factor $Z_c$, and composite $C_n$ are
\begin{equation}
 c_n=\min(n+1,H),\qquad Z_c=\sum_{i=0}^{c-1}a^i,\qquad
 C_n=\frac{1}{Z_{c_n}}\sum_{i=0}^{c_n-1}a^iF_{n-i}.
 \label{eq:target}
\end{equation}
We use $H=10$, $T=50$ ms, and $h=250$ ms, corresponding to captures at 20 Hz and a weight that halves after five capture intervals. The \textbf{temporal kernel} is the sequence of weights assigned to these image ages; here it is exponential and truncated after ten samples. During startup, $c_n$ grows with the number of acquired images, and normalization by $Z_{c_n}$ maintains unit total weight for the available samples. The newest capture contributes once to the sum, and the resulting composite remains displayed until the next scheduled capture. A continuously updated current-image overlay would require its own weighting rule because it would introduce an additional image between those captures. Resetting the effect clears the retained samples and restarts both the capture schedule and startup normalization.

Recursive accumulation provides a second formulation, $R_n=(1-a)F_n+aR_{n-1}$, in which every update combines the incoming image with the accumulated output. At arbitrary elapsed time $\Delta t$, retention $a(\Delta t)=\exp(-\lambda\Delta t)$, with $\lambda=-\ln(a)/T$, preserves the exponential decay constant. This adjustment controls the fading of available information, while obtaining a missing historical scene state requires a separate capture or reconstruction procedure. For example, an object moving between two delivered images may occupy an intermediate position at a scheduled capture time, and neither stored endpoint directly supplies that silhouette. Our native storage experiments remove this source of uncertainty by replaying independently rendered scene states at the exact target times. Hence, their results describe the storage and composition of supplied samples, with missed timestamps and visibility reconstruction reserved for a broader pipeline evaluation.

The offline reference evaluates \underline{Equation~\ref{eq:target}} in float64 arithmetic, providing a high-precision comparison for GPU composition. Inputs are captured before tone mapping, the transformation that maps scene color into a displayable range, so the temporal average is formed in the scene's linear color representation. The capture format is RGBA16F, comprising red, green, blue, and alpha channels stored as 16-bit floating-point values. Expanding those values to 32-bit floating point for replay preserves the captured values exactly, which allows a half-float history to retain the supplied source precision. A full-float source could contain additional precision or values beyond the half-float range, so the same preservation property depends on the chosen input format. Our image metrics use the three linear RGB channels, with alpha excluded from the error calculation.

\section{Storage Methods}
We evaluate three representations of the same scheduled image history. Dense storage retains every captured pixel, shared tiles combine image regions, and sparse innovations retain selected changes between captures. The following subsections explain how each representation recovers the target composite and accounts for its allocated storage.

\subsection{Dense Circular History and Color Precision}
The dense method stores $H$ complete images in a \textbf{circular buffer}, also called a \textbf{ring buffer}. This is a fixed collection of image slots with a write index that advances after each capture and wraps to the first slot after reaching the end. Once the history is full, the incoming image replaces the oldest image, and the index identifies which slots correspond to each retained age. A chronological array could provide the same logical history by shifting retained images or reorganizing references whenever a sample arrives. The circular arrangement allows storage to be reused through index arithmetic while preserving access to the complete set of samples. We combine those samples according to \underline{Equation~\ref{eq:target}} whenever a new capture is supplied.

One compute dispatch --- a submission of parallel GPU work --- both writes the incoming image and produces the weighted composite. Each invocation processes one pixel and owns the history locations for that pixel, allowing different invocations to update their data independently. With $P$ pixels, dense history requires $16PH$ bytes for RGBA32F, $8PH$ for RGBA16F, or $4PH$ for RGBA8, since the four channels occupy 16, 8, or 4 bytes per stored pixel. The packed implementations place four half-float channels in two 32-bit words or four 8-bit channels in one 32-bit word, then unpack them during composition. RGBA8 is an unsigned normalized format: values are clamped to $[0,1]$, rounded to one of 256 levels, and converted back to that interval when read. Every method also writes a common RGBA32F output image occupying $16P$ bytes, which is counted separately from the retained history.

\textbf{Memory totals} refer to the byte sizes of buffers explicitly allocated by the implementation, giving a reproducible account of history and workspace requirements. Physical residency describes the memory actually resident on the device, which may additionally depend on driver behavior, padding, and temporary storage generated by the shader compiler. Those quantities require separate instrumentation and remain outside the measurements reported here. For replay, the input sequence is preloaded before timing so that every representation receives the same images without file access or CPU image transfer during a measured update. This source cache is common experimental infrastructure and is excluded from the history totals. A live renderer would supply each incoming scene image directly, with the retained history and output allocations serving the roles described above.

\subsection{Synthesized Tiles}
The tile representation divides each image into rectangular $16\times16$ regions and stores their colors in a pool with a fixed number of available slots. Each historical timestamp contains references identifying the slots used by its tiles, allowing several timestamps to refer to the same stored region. When this capacity is exhausted, two retained tiles can be combined into a synthetic representative, and their references are redirected to the shared result. This operation approximates both tiles with one stored image region, so its quality depends on their color differences and their remaining contributions to future composites. Let $u_r$ and $v_r$ denote the collective normalized weights of those references at future output offset $r$, including the time at which each reference expires.
\clearpage 
\noindent For tile colors $A$ and $B$, minimizing squared future pair error for $X=\theta A+(1-\theta)B$ gives
\begin{equation}
 \alpha=\sum_r u_r^2,\quad \beta=\sum_r v_r^2,\quad
 \gamma=\sum_r u_rv_r,\quad
 \theta=\frac{\alpha+\gamma}{\alpha+\beta+2\gamma}.
\end{equation}
For $G=\alpha+\beta+2\gamma>0$, the incremental pair cost is
\begin{equation}
 E_{AB}=\frac{\alpha\beta-\gamma^2}{G}\operatorname{MSE}_{RGB}(A-B).
\end{equation}
Here, $\operatorname{MSE}_{RGB}$ is the mean squared difference over the three color channels and all pixels in a tile. The implementation evaluates the equivalent squared residual profile to reduce numerical cancellation when the terms $\alpha\beta$ and $\gamma^2$ are close. Startup normalization and finite expiration are included in the temporal profiles, ensuring that a tile's priority reflects its contribution over the remaining history window. A pair whose references have all expired has zero future weight and can be released. For each candidate pair, the least-squares representative minimizes the stated pair objective; repeated greedy merging retains approximation errors from earlier decisions. Cross terms involving those discarded originals are omitted, giving a local decision rule whose cumulative error is evaluated experimentally.

Five dependent GPU stages perform candidate preparation, global priority updates, retained-value copying, incoming-data admission, and final image composition. A heap is the priority data structure used to locate the least costly merge; its ordering allows the minimum to be found without scanning every candidate for each decision. Several references may identify the same slot (a condition known as \textit{aliasing}), so retained values are copied before a reused slot can overwrite information needed by another reference. Dispatch ordering and synchronization make those writes available to subsequent stages, and the memory accounting includes their actual retained storage and workspace. An ablation (i.e., a comparison with one component removed) sets the profile interaction term $\gamma$ to zero to examine the role of joint temporal weighting. We also evaluate a packed-color variant to examine precision and allocation together, using explicitly stated dense formats as comparison points.

\subsection{Sparse Innovations}
The \textbf{sparse innovation} representation stores selected changes between successive images. An \textit{innovation} is a signed pixel difference, $\Delta_n=F_n-F_{n-1}$, containing positive components where a channel becomes brighter and negative components where it becomes darker. This permits unchanged pixels to be represented through the current image and the retained changes associated with that pixel. The term \textit{sparse} refers to allocating entries for selected changes across the image, in contrast to reserving every pixel at every historical timestamp. Rewriting \underline{Equation~\ref{eq:target}} in terms of these differences gives
\begin{equation}
 C_n=F_n-\sum_{j=0}^{c_n-2}b_j\Delta_{n-j},\qquad
 b_j=\frac{\sum_{i=j+1}^{c_n-1}a^i}{Z_{c_n}}.
 \label{eq:innovation}
\end{equation}
The coefficient $b_j$ gives the total historical weight affected by the change of age $j$, allowing composition to begin with the exact current image and subtract weighted changes. Each retained event records a pixel address, the capture at which it was created, and its signed color difference. Its birth time determines both the current coefficient and the capture at which its contribution leaves the finite window. Omitting an event changes the reconstructed composite during that remaining interval, with its effect disappearing after expiration. Signed contributions can yield reconstructed values outside the range of the supplied colors, so image metrics are evaluated on the unclamped output. This preserves overshoot as part of the measured approximation error.

We assign each event a priority equal to its squared RGB magnitude multiplied by the sum of its squared remaining coefficients. The coefficients include startup normalization and finite expiration, making the priority an estimate of the event's isolated contribution to future squared error. When the pool is full, the lowest-priority event is discarded as required to admit a more valuable incoming change. Events are also organized into lists associated with each pixel, allowing reconstruction to visit the retained changes that affect that location. The score treats events independently, with cancellation or reinforcement between signed changes contributing additional cross terms to the measured reconstruction error. The logical update below is implemented through dependent GPU dispatches for classification, serial heap and list updates, and final composition.
\begin{lstlisting}[language=Python]
expire_events_outside_window()
changes = incoming_color - previous_color
for event in nonzero_pixel_changes(changes):
    weight = future_coefficient_energy(event)
    score = weight * rgb_energy(event)
    admit_or_replace_minimum(event, score, capacity)
rebuild_pixel_lists()
output = incoming_color - weighted_retained_changes()
previous_color = incoming_color
\end{lstlisting}
For capacity $S$, explicit history and workspace allocation is
\begin{equation}
 M(P,S)=24P+36S+16\quad\text{bytes}.
 \label{eq:memory}
\end{equation}
The $24P$ term includes previous color, per-pixel list heads, and priorities computed for incoming changes. The $36S$ term includes event color, pixel and birth metadata, heap indices, links between list entries, and retained priorities. Sixteen control bytes complete the allocation, with the common output image counted separately as in the dense methods. These auxiliary values are needed to identify which changes belong to each pixel and determine when entries can be reused. The available event capacity therefore follows from the complete memory expression, since counting event colors alone leaves out storage required by admission and reconstruction.

The dense method performs $O(PH)$ work because every output pixel reads up to $H$ retained colors. Sparse classification and composition take $O(P+S)$ work, with $O(K\log S)$ additional work for $K$ heap updates. The heap updates are serialized in this implementation, causing their execution time to depend on a centralized sequence of allocation decisions. Tile merging has related costs from classifying candidates, maintaining priorities, and updating shared references. These operations explain why reducing the number of stored color values can increase measured GPU duration: parallel access to a regular dense array and coordinated access to an adaptive pool use the device differently.

\section{Experimental Procedure}
The evaluation separates reconstruction accuracy from GPU timing so that image readback and scene capture can be performed independently of measured storage updates. Development streams examine the adaptive methods at low resolution; subsequent experiments compare dense precision formats at 720p and 1080p. We describe the scene inputs, timing protocol, and error definitions below.

\subsection{Development and Native Follow-up}
Development experiments use six recorded motion streams at $160\times90$, covering translation, rotation, crossings, occlusion, camera motion, and a private deforming model (Link's model, ripped from \emph{Majora's Mask}). Translation changes image position with a comparatively simple silhouette, while rotation and deformation also change the object's projected shape. Crossings and occlusion examine overlapping trails and the visibility changes produced when one object passes in front of another. These streams provide controlled cases for examining approximation errors and allocation behavior during algorithm development. The public source package excludes the private model and its captures, and \underline{Figure~\ref{fig:failure}} uses the procedural crossing scene.

The native-resolution follow-up uses two procedural perspective scenes, containing a rotating asymmetric object and a moving camera. Each scene is rendered independently at 16 exact target times from 0 to 0.75 s, at $1280\times720$ and $1920\times1080$, producing four input streams. Evaluating scene state as a function of absolute time ensures that a given timestamp corresponds to the same object and camera configuration across storage methods. \textbf{Temporal antialiasing (TAA)} and \textbf{multisample antialiasing (MSAA)} are disabled to avoid introducing additional temporal reuse or sample averaging into the captured inputs. Materials are unshaded, and capture waits two render frames after setting each state so that the renderer has processed the requested configuration. This design isolates the history representation under controlled motion and provides a reproducible basis for the precision comparison.

Inputs were captured with Godot 4.8 beta3, and the native precision and timing follow-up uses pinned Godot 4.7.2 stable. GPU replay runs through the Vulkan Forward+ renderer on an NVIDIA GeForce RTX 3060 Ti with driver 610.88 under Windows 11. The captured scene images are supplied to each representation in the same order, with identical history and weighting parameters. Measuring replay separates the storage update and composition from the cost of generating scene geometry, shading, and capture files. Engine hashes, configurations, and source run identifiers are retained with the experiment records to identify the software and data used for each comparison.

\subsection{GPU Timing Protocol}
Each dense timing configuration has five independent process runs, so a repetition includes a fresh execution of the experiment. Every run warms up for at least 120 frames and 2 s, then measures at least 300 frames and 10 s while repeating complete 16-capture cycles with resets. Vsync is disabled, and four methods, including a no-effect control, rotate through five order blocks to vary their measurement positions. This rotation provides partial order balancing, with some positions represented more often than others. GPU timestamps surround the storage work, and their asynchronous results are collected after completion without waiting on each measured frame. Accuracy readback, which transfers output images to the CPU for comparison, is performed in separate runs; the no-effect timestamps resolve to zero at the available granularity and are retained as a below-resolution control.

\subsubsection{Aggregation}
We aggregate timing at the process-run level because frames within one execution are temporally correlated. The central duration is the median of five run medians, accompanied by the observed range of those medians. Paired comparisons match logical order blocks, with some repetitions collected in a later batch approximately 15 hours after the initial measurements. Tail measurements summarize each run's percentiles before taking their median across runs; for example, p99 is the duration at or below which 99\% of that run's measured updates fall. These statistics describe the storage and composition pass, excluding scene rendering, source preload, and output readback. Total GPU frame time, complete-application frame rate, physical residency, and hardware-counter bandwidth would require additional measurements and remain outside this comparison.

\subsection{Image Fidelity Metrics}
We measure image error against the float64 offline reference, also termed the oracle, which computes the specified target directly from the supplied scene images. \textbf{Mean squared error (MSE)} averages the squared difference between corresponding RGB values, and \textbf{root mean square (RMS)} error is its square root. We evaluate both across the full image and within a motion region containing every pixel whose supplied color changes from the first image by more than $10^{-6}$, including pixels occupied only by earlier silhouettes. \textbf{Peak signal-to-noise ratio (PSNR)} expresses MSE on a logarithmic scale with data range 1, with larger values indicating smaller numerical error. Selected final frames also use the \textbf{structural similarity index (SSIM)} \citep{Wang2004SSIM}, computed on luminance with Rec.~709 coefficients, Gaussian $\sigma=1.5$, an 11-pixel window, population covariance, $K_1=0.01$, $K_2=0.03$, and a five-pixel border excluded. These measures characterize numerical and structural similarity under the stated definitions, with perceived visual equivalence requiring a separate human evaluation.

\section{Results}
We first report the native-resolution precision comparison, using the same supplied scene images for each dense format. We then examine the adaptive representations through their allocated memory, reconstruction error, and measured execution time. The repeated timing results and preliminary feasibility measurements are identified separately throughout this section.

\subsection{Dense Precision at Native Resolution}
\begin{figure}[htbp]\centering
\includegraphics[width=\linewidth]{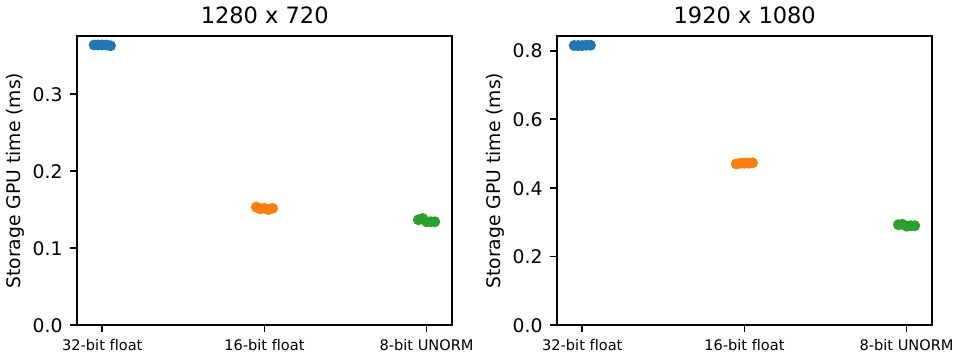}
\caption{Native-resolution dense storage timing. Points are independent process-run medians. The measurements concern storage and composition, excluding scene rendering.}\label{fig:timing}
\end{figure}

\underline{Table~\ref{tab:timing}} and \underline{Figure~\ref{fig:timing}} report the effect of dense storage precision on allocation and GPU duration. Relative to RGBA32F, RGBA16F halves history storage and has median paired speed ratios of 2.399 at 720p and 1.727 at 1080p. A speed ratio divides the RGBA32F duration by the comparison duration within each logical block, after which we take the median of the paired values. RGBA8 quarters history storage and produces corresponding ratios of 2.702 and 2.816. The table summarizes each format's run statistics independently, so the ratio of two table medians may differ from the paired statistic. Together, the allocation and duration measurements describe the practical benefit of packing the same retained image history into fewer bytes.

\begin{table}[htbp]
\centering\small
\begin{tabular}{llrrrr}
\toprule
Size & History & MiB & Median ms & Range ms & p99 ms\\\midrule
720p & RGBA32F & 140.63 & 0.3635 & 0.3625--0.3635 & 0.6538\\
 & RGBA16F & 70.31 & 0.1516 & 0.1500--0.1531 & 0.1690\\
 & RGBA8 & 35.16 & 0.1341 & 0.1341--0.1382 & 0.1536\\\midrule
1080p & RGBA32F & 316.41 & 0.8151 & 0.8151--0.8161 & 1.1442\\
 & RGBA16F & 158.20 & 0.4721 & 0.4700--0.4731 & 0.7522\\
 & RGBA8 & 79.10 & 0.2898 & 0.2883--0.2939 & 0.3226\\\bottomrule
\end{tabular}
\caption{Dense storage results, five process runs per row. Range refers to run medians. The p99 column is the median run-level p99. MiB counts history only; common output allocation is additional.}\label{tab:timing}
\end{table}

All 192 native output comparisons passed the format-specific reference tolerance of $3\times10^{-6}$ maximum channel error. This test compares the GPU output with a reference that applies the same storage conversion before composition, separating implementation correctness from the error introduced by reducing precision. Against the unquantized oracle, RGBA32F and RGBA16F have identical measured full-frame MSE of approximately $3.3$--$3.5\times10^{-17}$. Their agreement follows from\linebreak the source format: every captured half-float value can be represented by either history, leaving only the numerical effects of composition. The result therefore supports using half-float history for these inputs and explains why the corresponding error map appeared black at the RGBA8 display scale. \underline{Figure~\ref{fig:teaser}} reports the half-float error numerically and reserves the error image for the quantized output.

\begin{table}[htbp]\centering\small
\begin{tabular}{llrrr}\toprule
Scene & Size & Full RGB MSE & Motion RGB MSE & PSNR dB\\\midrule
Rotation & 720p & $1.389\times10^{-6}$ & $1.120\times10^{-6}$ & 58.57\\
Camera & 720p & $1.196\times10^{-6}$ & $8.496\times10^{-7}$ & 59.22\\
Rotation & 1080p & $1.389\times10^{-6}$ & $1.120\times10^{-6}$ & 58.57\\
Camera & 1080p & $1.196\times10^{-6}$ & $8.491\times10^{-7}$ & 59.22\\\bottomrule
\end{tabular}
\caption{RGBA8 error against the unquantized float64 oracle, averaged across the 16 outputs in each stream. Format-specific correctness and quantization fidelity are separate comparisons.}\label{tab:precision}
\end{table}

\clearpage
The RGBA8 results in \underline{Table~\ref{tab:precision}} quantify the effect of rounding scene color to 8-bit normalized channel values. Selected final-frame SSIM at 1080p is 0.99419 for rotation and 0.99630 for camera motion, complementing the full-frame and motion-region MSE measurements. A separate 90-output edge suite covers random \textbf{high-dynamic-range (HDR)} values in $[-2,8]$, zeros and ones, odd $17\times9$ images, histories of length 1 and 10, resets, and resource recreation between cases. Its maximum format-reference error is $1.10\times10^{-6}$, below the declared $5\times10^{-6}$ tolerance. RGBA8 is compared with its explicitly clamped and quantized reference in that suite, making the tolerance a check of the specified conversion, including its handling of out-of-range HDR values. The correctness check and the fidelity measurements consequently assess their respective questions using the same storage rule.

\subsection{Adaptive Storage and Its Costs}
At $160\times90$, float tile synthesis allocates 373,084 bytes for 75 slots, while an endpoint-retention comparison fits 82 slots in 372,744 bytes. Endpoint retention represents a selected pair using one of its existing tile images, providing a comparison for the benefit of constructing a blended representative. Synthetic representatives improve selected comparisons against that rule, and a 142,684-byte packed synthesis variant produces 6.9--91.3 times lower motion-region MSE than a 144,000-byte RGB332 dense control across the six development streams. RGB332 assigns three bits to red, three to green, and two to blue, totaling eight color bits per pixel; RGBA8 assigns eight bits to each of four channels, totaling 32 bits per pixel. A 288,000-byte RGB565 dense representation, using five red, six green, and five blue bits, dominates the float synthesis candidate on two later scenes. Hence, these comparisons show how the apparent benefit of synthesized representatives depends on the precision and memory budget of the dense alternative.

\begin{figure}[htbp]\centering
\includegraphics[width=\linewidth]{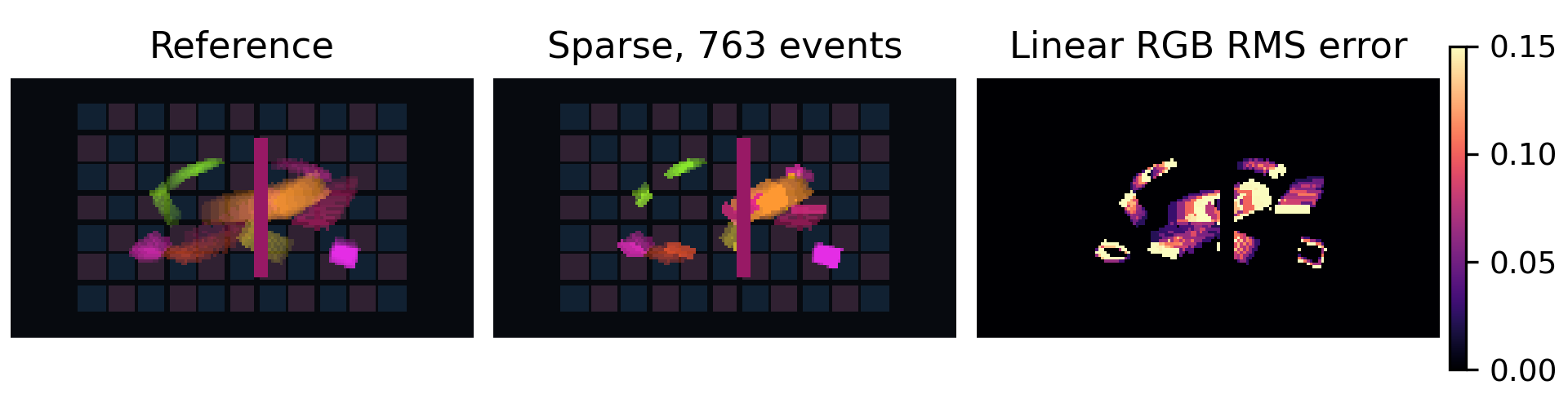}
\caption{Procedural crossing example at $160\times90$. With 763 retained events, the sparse representation loses changes needed by the finite history. The error image uses a fixed linear-RGB scale.}\label{fig:failure}
\end{figure}

The innovation experiments also illustrate the importance of including workspace in a storage budget. An initial projection allowed 5,944 events within a nominal 373,084-byte allocation by leaving out required workspace. \underline{Equation~\ref{eq:memory}} gives an actual requirement of 559,600 bytes for that capacity, reducing the number of events that fit the full budget to 763. At the larger capacity, five of six development streams had near-zero error; at 763 events, rotation and occlusion retain that behavior. Float tile synthesis produces lower error in four streams under the corrected comparison, and \underline{Figure~\ref{fig:failure}} illustrates the loss of historical changes in a crossing sequence. The corrected accounting is therefore part of the fidelity result, since it changes how much information the method can retain at the declared memory limit.

A repeated low-resolution timing comparison uses five processes per method on the same recorded 40-capture camera stream. Dense storage has a median duration of 0.008192 ms, and sparse innovations have 3.647488 ms, with median p95 of 4.554752 ms and p99 of 4.975704 ms. Sparse history and workspace occupy 373,084 bytes compared with 2,304,000 bytes for dense history, reducing the counted allocation by 83.81\%. The median paired duration ratio is 445.25, indicating the scale of the GPU cost associated with that reduction in this implementation. These development timings use the beta engine and form their own repeated comparison, with the stable-engine native results reported separately in \underline{Table~\ref{tab:timing}}.

Before scheduling long native sparse runs, we applied a 50 ms feasibility budget corresponding to the 20 Hz capture interval. A pilot is a short preliminary run used here to determine whether the implementation can complete its storage update within that interval. Sixteen-frame pilots measured median storage durations of 386.68 ms at 720p and 900.72 ms at 1080p, with history and workspace allocations of 22.77 and 51.24 MiB. Both exceed the budget, triggering the declared decision to stop before repeated native sparse timing. Their purpose and protocol differ from the warmed, five-run measurements in \underline{Table~\ref{tab:timing}}, and their values are interpreted as feasibility observations. They identify the current implementation as too costly for the selected capture schedule while motivating investigation of its allocation and scheduling structure.

\section{Discussion}
The dense precision results follow from the relationship between source representation and history representation. A half-float history stores every supplied scene value and retains the regular per-pixel access pattern of the full-float implementation. Adaptive storage introduces additional operations to identify reusable information, maintain addresses and priorities, and reconstruct values through indirect references. In the sparse implementation, the serial heap also requires a centralized sequence of decisions as changes arrive across the image. Camera motion can generate changes over much of the scene, increasing the work that passes through this allocation stage. These observations motivate a future comparison of distributed admission decisions and simpler local policies, with any resulting speed improvement requiring new measurements against the packed dense controls.

\subsection{Finite History and Reconstruction}
Finite expiration also explains part of the difficulty of reducing temporal state exactly. Consider the mapping from $H$ old samples to their next $H$ outputs after subtracting the contributions of known future inputs. For kernel weight $k_i$ and output normalization $Z_r$, the coefficient matrix is $W_{ri}=k_{i+r}/Z_r$ when $i+r<H$, and zero otherwise. Reversing its columns makes it triangular, with a nonzero diagonal when the final kernel weight $k_{H-1}>0$, so the matrix has rank $H$. Rank counts the independent components of this linear mapping, giving a requirement of $H$ temporal components for universally exact linear reconstruction. This places the practical opportunities in nonlinear compression, scene-dependent redundancy, spatial sharing, and controlled approximation, while the finite removal of old samples distinguishes the target from an indefinitely supported exponential recurrence.

We present this work as an empirical study of representation choices and their implementation costs. The least-squares tile objective and signed-change representation connect to established approximation and reconstruction ideas, and the experiments provide evidence about the specific methods evaluated here. Their comparison with packed dense history gives a practical reference for assessing both successful compression and its reconstruction overhead. Residual-summary experiments and alternate scheduling variants are retained in the accompanying code and evidence, preserving the development cases that informed this comparison. Algorithmic novelty remains unestablished, and the supported outcome is the measured behavior of these representations under their stated fidelity and resource constraints.

\subsection{Limitations}

The measured scope comprises one GPU, two short native scenes, one history length, and one temporal kernel, with controlled unshaded inputs. Generalization to transparency, postprocessing, TAA, upscaling, and frame generation requires experiments that include their interactions with retained history. Exact replay also isolates the representation from capture scheduling and reconstruction errors, leaving the cost and fidelity of a complete motion-assisted pipeline for subsequent evaluation. Perceptual assessment would require a separate study, and quantitative historical validation would require timestamped reference captures with a documented game version and rendering environment; the illustrative images in \underline{Figure~\ref{fig:mm-reference}} have unspecified capture conditions. The 20 Hz history schedule serves as an experimental parameter throughout this paper, with regional or version-specific Nintendo 64 behavior requiring independent historical verification.

\clearpage
\section{Conclusion}

\vspace{-0.25em}
In this work, we evaluated the storage and GPU composition of a finite sequence of fading historical images. Dense half-float storage reduced both allocated history memory and measured GPU duration while preserving the precision of the supplied scene inputs. RGBA8 provided further savings through quantization, with its image error measured against the same independently composited reference. Synthesized tiles and sparse innovations reduced stored information through conditional reuse and selective retention, and their reconstruction and allocation costs exceeded those of the practical dense alternatives in the evaluated comparisons. These results support matching history precision to source precision and accounting for the complete workspace when comparing adaptive representations. Subsequent work can use the supplied implementations and reference calculations to evaluate decentralized allocation, broader scene content, and additional GPU architectures under the same fidelity and resource definitions.

\vspace{-0.4em}
\section*{Index of Supplemental Materials}

\vspace{-0.25em}
The source code is available at \href{https://github.com/danielathome19/stylized-temporal-persistence}{github.com/danielathome19/stylized-temporal-persistence} under the \textbf{MIT License}. The repository includes the Godot compute implementations, procedural scenes, experiment configurations, independent reference calculations, and relevant correctness tests. Its README describes environment setup and the commands for capturing scene inputs, comparing precision, recording GPU timings, and generating reports and experimental figures. Raw experiment outputs are maintained separately from the source package.
Private model assets are excluded from distribution, and the MIT License applies to the project source code; third-party assets and dependencies retain their respective ownership and licensing terms.

\vspace{0.5in}
\begin{center}
\begin{minipage}{\linewidth}
\centering
\includegraphics[width=0.485\linewidth]{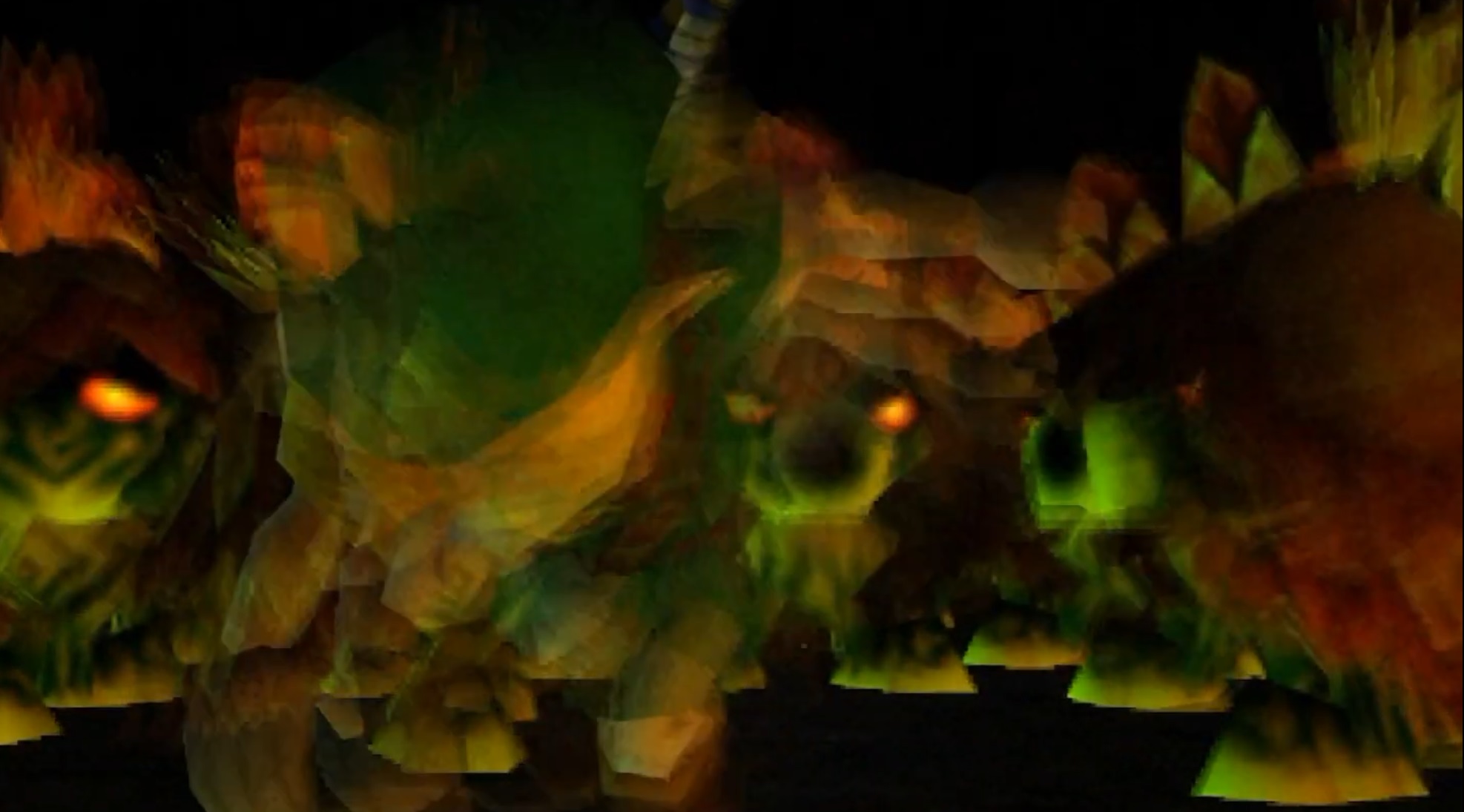}\hfill
\includegraphics[width=0.485\linewidth]{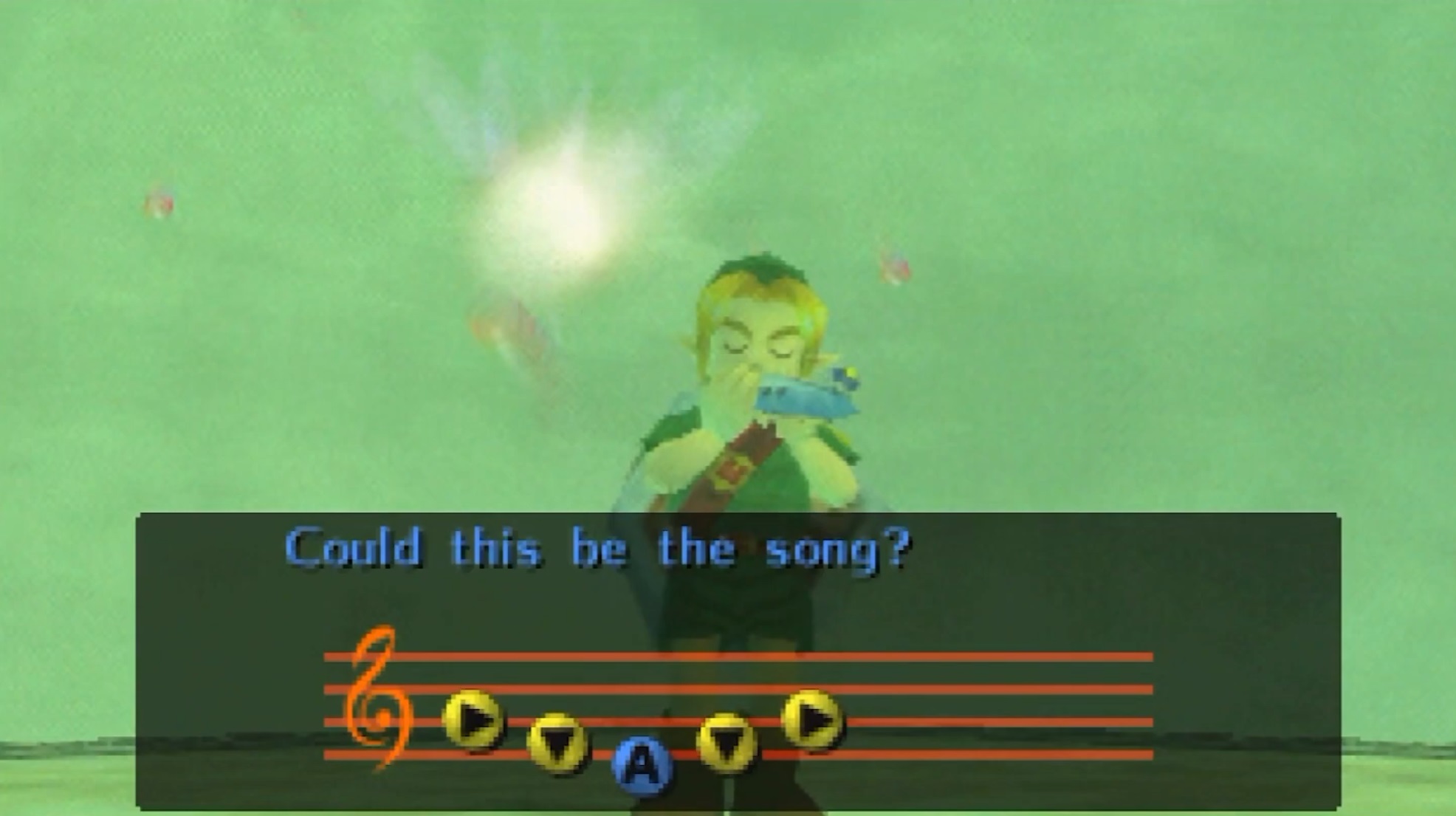}
\captionof{figure}{Additional examples of temporal persistence in \emph{The Legend of Zelda: Majora's Mask}: overlapping scene silhouettes (left) and fading/trailing scene imagery behind the interface (right). Game imagery \copyright\ Nintendo.}
\label{fig:mm-additional}
\end{minipage}
\end{center}

\clearpage
\bibliographystyle{jcgt}
\bibliography{references}

\section*{Author Contact Information}
Daniel Szelogowski\\
University of Wisconsin--Whitewater\\
Department of Computer Science\\
Whitewater, WI 53190\\
\href{mailto:szelogowdj19@uww.edu}{szelogowdj19@uww.edu}
\end{document}